# Design of new superconducting materials, and point contact spectroscopy as a probe of strong electron correlations

Laura H. Greene, Hamood Z. Arham, and Cassandra R. Hunt, and Wan Kyu Park

Center for Emergent Superconductivity
Department of Physics and Frederick Seitz Materials Research Laboratory, University of Illinois at Urbana-Champaign, Urbana, IL 61801 USA

*ABSTRACT:* At this centenary of the discovery of superconductivity, the design of new and more useful superconductors remains as enigmatic as ever. These materials play crucial roles both for fundamental science and applications, and they hold great promise in addressing our global energy challenge. The recent discovery of a new class of high-temperature superconductors has made the community more enthusiastic than ever about finding new superconductors. Historically, these discoveries were almost completely guided by serendipity, and now, researchers in the field have grown into an enthusiastic global network to find a way, together, to predictively design new superconductors. After a short history of discoveries of superconducting materials, we share our own guidelines for searching for high-temperature superconductors. Finally, we show how point contact spectroscopy (PCS) is used to detect strong correlations in the normal state with a focus on the strongly-correlated region in the normal state of the Fe-based superconductors, defining a new region in the phase diagram of $Ba(Fe_{1-x}Co_x)_2As_2$.

At this ten carat diamond anniversary (100 years) of the discovery of superconductivity, the silver anniversary (25 years) of the discovery of high-temperature superconductivity, and the blue topaz anniversary (4 years) of the discovery of a new class of high-temperature Fe-based superconductors, it is worth looking back at the history, and modes of discovery of the many classes of known superconductors.  We do this not only for historical breadth, but also to see if we can learn to turn our serendipitous discoveries into predictive design of new superconductors.  We start with a short history of the field, mention the ubiquitous connection between the tunable superconductors we know of today, their phase diagrams, and demonstrate how we can measure strong correlations in these materials with point contact spectroscopy (PCS).

The Physics Today article by van Delft and Kess [1] beautifully relates the discovery of superconductivity by Heike Kamerlingh Onnes:  The quest for achieving lower temperatures was followed by the curiosity-driven measurements of the resistance of metals at these newly achieved liquid helium temperatures.  Mercury was chosen since it was a liquid at room temperature, and therefore easily purified.  As described by Stephen Blundell [2], there existed three popular predictions on how the resistance would behave upon approaching absolute zero:  Lord Kelvin predicted an insulating up-turn at low temperatures due to a freezing out of charge carriers; Matthiessen thought the resistance would decrease to a finite value; and Dewar predicted a smooth decrease to zero resistance at zero temperature.  The abrupt drop in resistance at ~ 4 K was a complete surprise!  Over the next few decades, the critical temperature slowly increased through systematic tests of elements, alloys and compounds. From the early 1950s, this effort was primarily led by Bernd Matthias, and in 1952, Matthias discovered the first "new class" of superconductors by combining ferromagnetic and semiconducting elements: $CoSi_2$, with a $T_c$ of 1.33 K [3]. Much of this history is described in a lovely monograph [4], which includes "Matthias' Rules" for finding new superconductors:

1. Transition metals are better than simple metals;
2. Peaks of density of states at the Fermi level are good;
3. High crystallographic symmetry is good: Cubic is best;
4. Stay away from oxygen, magnetism, and insulating phases.

These rules were very helpful for designing new conventional, metallic superconductors, but clearly did not lead us to discover new families of unconventional and high-temperature superconductors.

Nearly concomitant with Matthias' discovery of the first new class of superconductors was the discovery by John Hulm and George Hardy of the first of the "A15" family of superconductors [5]. These compounds have the $A_3B$ structure with A being a transition metal. This discovery was particularly significant for applications as these superconductors were the first to exhibit high critical currents in the presence of strong applied magnetic fields. In the following years, Bernd Matthias and colleagues went on to discover over thirty A15 compounds with values of $T_c$ reaching ~ 23 K for $Nb_3Ge$ and at low temperatures, could exhibit a critical field of ~ 37 T [6]. Another crucial discovery occurred in the early 1960's, when John Hulm at Westinghouse in the US, was able to make practical wires out of a superconducting random alloy discovered at the Rutherford Appleton Labs in the United Kingdom: Nb:Ti [7]. These wires exhibit a $T_c$ ~ 9.5 K and can remain superconducting in fields as high as 14 T. Note that the $T_c$ and critical magnetic fields are not as high as the A15s, but by virtue of Nb:Ti being a random alloy, as opposed to a compound, this material is significantly more malleable and robust, and therefore more highly suited for applications. Applications for Nb:Ti superconducting wires today include magnets in the Large Hadron Collider (LHC), the ITER project for controlled nuclear fusion located in France, and in MRI machines used for diagnosis in hospitals. At the retirement party for Jack Wernick, who also discovered many A15 compounds, John Rowell stated in his honor, "High $T_c$ gets Nobel prizes and high $J_c$ saves lives." Although a good part of this history and the fundamental research today is in searching for new families of higher-temperature superconductors, even room temperature superconductors, it is worth stressing that there is no guarantee that such a material will have any practical applications at all. Our understanding of the fundamental limits of the current carrying capabilities in superconductors, and their control, remains a grand challenge.

In 1979, Frank Steglich discovered a completely new class of superconductors: The heavy fermion superconductors [8]. Heavy fermion materials have an antiferromagnetic ground state with either rare earth ions (4f configurations) or actinides (5f configurations) as part of the crystal lattice. The Kondo scattering of the itinerant electrons off the localized f-electrons becomes coherent at low temperatures, and the hybridization of the conduction electrons with the f-electrons leads to extremely large electron effective masses; tens to a thousand times the free electron mass [9]. This discovery was particularly important for laying the foundation for high-$T_c$ materials, as they were the first truly tunable superconductors through the competition between ground states. Also, for the first time it appeared that magnetism looked beneficial for achieving a high-$T_c$, and the electron-phonon mechanism in the BCS theory was breaking down. The study of heavy fermion superconductors has turned out to be extremely fruitful for materials physics. First, the discovery of the "115 family" (CeMIn$_5$, M=Co,Rh,Ir) was not completely serendipitous, but driven by guidelines acquired from many studies of existing superconductors, and second, it was shown that these novel and new heavy fermion superconductors were quantum critical [10,11]. The ubiquitousness of the phase diagram with competing phases and the emergent superconducting dome was becoming clearer.

Parallel to these studies on metals were those in the search for oxide superconductors. Here lie a few more examples of predictive discovery of superconductivity. In 1964, Marvin Cohen theoretically predicted the first semiconducting superconductor [12], the oxide SrTiO$_3$, which easily superconducts with cation doping or oxygen reduction. In 1983, Len Mattheiss and Don Hamann performed electronic structure calculations [13] on the existing superconductor BaPb$_{1-x}$Bi$_x$O$_3$, which later led Mattheiss, a theorist, to go into the lab and grow the new Ba$_{1-x}$K$_x$BiO$_3$ superconductor [14], then refined by Bob

Cava and co-workers [15] to realize the high $T_c$ of nearly 30 K. In the middle of this, in 1986, George Bednorz and Alex Muller made the truly transformative discovery of the high-$T_c$ superconductivity in the cuprate La-Ba-Cu-O [16]. The subsequent discovery of superconductivity above the temperature of liquid nitrogen by Paul Chu and co-workers [17] was soon followed by the record high $T_c$ of 165 K in the $HgBa_2Ca_2Cu_3O_{8-x}$ system under pressure [18] which is still held today.

In 2006, Basic Energy Sciences of the Department of Energy published a report [19], which contained a comprehensive history of the field, its present status, and what might be needed to invent new high-temperature superconductors. Besides an obligatory $T_c$ vs. time plot, the canonical phase diagram was included. The universal quantum criticality of the tunable superconducting materials, including the Femi liquid behavior on the over-doped side, the "strange metal" on the underdoped side, and the emergent superconducting dome arising above the quantum critical point, was established. The astonishing similarity to other tunable superconductors, including heavy-fermions and organics, was noted. At the time it was conjectured that this phase diagram would act as a template for new "Matthias's Rules" for high-$T_c$. But, in the 22 years since the discovery of the cuprates, no new families of tunable high-$T_c$ superconductors were discovered.

Enter Hideo Hosono, a brilliant and dynamic materials chemist from the University of Tokyo. While studying, among many other things, transparent conductors, he discovered the Fe-based superconductors [20]. Although this breakthrough was in many ways serendipitous, Hosono knew immediately what he had found and soon caused the transition from the "copper age" to the "Iron age" [21] in high-temperature superconductivity. Zhongxian Zhou and co-workers at the Chinese Academy of Sciences in Beijing systematically tested a variety of substitutions and dopings, and pushed the high-$T_c$ Fe-based materials to $T_c$ = 58 K which remains the limit today [22].

There exists an international alliance, supported by I2CAM (International Institute for Complex and Adaptive Matter) chaired by Rick Greene at and the first author (no relation), with this charge:

> *There is now strong international support to find new classes of "better" superconductors. For any one of us, putting all of our efforts toward discovering new superconductors is too risky. We are therefore sharing our expertise and resources, on a worldwide scale, to search for new classes of superconductors.*

We have a working group, open to all [23].

As part of our I2CAM charge, we at the CES (Center for Emergent Superconductivity), readily share our six principles to guide our search for new superconductors [24}:

1. Reduced Dimensionality
2. Transition metal & other large Hubbard U ions
3. Light atoms
4. Tunability

5. Charged and multivalent ions
6. Low dielectric constant.

Although some of these principles are inconsistent with each other, they are motivated by past studies with some theoretical inspirations. They are meant to help us design materials with competing phases, leading to emergent phases. The goal is to design the emergent phase to be a superconductor, which remains a grand challenge [25].

In order to understand the ubiquitous phase diagrams of the tunable superconductors, we have undertaken point contact spectroscopy (PCS) to understand the role and existence of strong electron correlations on the "strange metal" or "left side" of the phase diagram.

PCS is accomplished when current-voltage characteristics are obtained across a small metallic contact, with radius less than the mean free path of the material. In that case, quasiparticle energy gain or loss mostly occurs near the contact, with the length scale determined by the quasiparticle scattering length. Nonlinearities in the conductance characteristics then reflect energy-dependent quasiparticle scattering near the contact region. There are many good reviews on this technique [26,27,28,29].

Over the years, PCS has most successfully been applied to superconducting materials, where the details of the energy gap and electron-phonon coupling can be easily extracted from the conductance data according to the BTK model [30]. Although the conductance spectra derived from Andreev reflection across the interface look completely different from those obtained by quasiparticle tunneling spectroscopy, correct analysis yields the same results, including superconducting gap size, order parameter symmetry, and in conventional superconductors, the electron-phonon interaction, $\alpha^2(\omega)F(\omega)$ [31,32].  Beyond this well-understood spectroscopy for single-band superconductors, this technique can detect a whole host of strong electron correlations.  In fact, the original application was to measure phonon spectra in normal metals:  In applying a bias across a small aperture in the ballistic limit, a non-equilibrium quasiparticle occupation is induced through the junction according to the Boltzmann equation.  The bias dependence of the conductance depends on the probability of quasiparticle scattering (backflow through the junction) and that can be inelastic energy loss, as in the case of phonons, or elastic, as in the case of many-body effects such as superconductivity (Andreev reflection) and Kondo scattering.

 Our group has been successful in applying this technique to the normal state in strongly – correlated metals:  We have detected the onset and growth of the Kondo lattice in the "115" heavy fermions [28,31]; observed the onset of antiferromagnetic ordering in strongly-correlated spin density wave systems [33,34]; and directly measured the hybridization gap and Fano resonance in $URu_2Si_2$ [35].  Here we present results in which strong electron correlations are detected at high-temperature in the Fe-based superconductors [36].

PCS on the parent and doped series of the "122" Fe-based superconductor Ba(Fe$_{1-x}$Co$_x$As)$_2$, reveals a strong "gap-like feature" (GLF) at temperatures well above the antiferromagnetic and structural transitions (Fig. 1). As explained above, this detected GLF must be due to strong electron correlations. We have measured over 100 junctions in this series and were able to define a new line in the phase diagram, as shown in Fig. 2. Reproducible GLF-type features have also been obtained in the SrFe$_2$As$_2$ and Fe-chalcogenides. We do not have an explanation for these reproducible and robust GLFs, and do not even claim this conductance enhancement is due to a gap, but note there are other groups who see interesting effects with onset temperatures well above $T_N$, comparable to the temperatures at which we detect the GLF. In crystals that are gently detwinned by application of uniaxial stress, electronic orthorhombicity is measured in temperature-dependence of the resistivity [37,38], far-infrared spectroscopy [38], and ARPES [39]. In untwined crystals, electronic nematicity is reported by inelastic neutron scattering [41], STM [42] and Matsuda has reported a C4 symmetry breaking in BaFe$_2$(P$_x$As$_{1-x}$)$_2$ [43]. Our conjecture is that all of these experiments are related, that we are detecting orbital ordering, but the exact origin of our GLF remains a mystery.

In conclusion, we have learned a great deal from the eminent scientists who have made transformative discoveries in superconducting materials, weather their discoveries have been incremental, serendipitous, or predictive. We have also shown how point contact spectroscopy can be an important tool for mapping out and understanding the seemingly universal phase diagrams of tunable superconductors. It is important to continue to learn from each other, and work together to understand the mechanisms of our existing superconductors, and learn, together how to make "better" ones.

Acknowledgements:  The PCS work has been in collaboration with P.H. Tobash, F. Ronning, E.D. Bauer, J.L. Sarrao, J.D. Thompson, J. Gillett, S.D. Das, S. E. Sebastian, A. Thaler, S.L. Bu'dko, P. C. Canfield, Z. J. Xu, J. S. Wen, Z. W. Lin, Q. Li, and G. Gu.  This work is supported by the Center for Emergent Superconductivity, an Energy Frontier Research Center funded by the US DOE Award No. DE-AC0298CH1088.

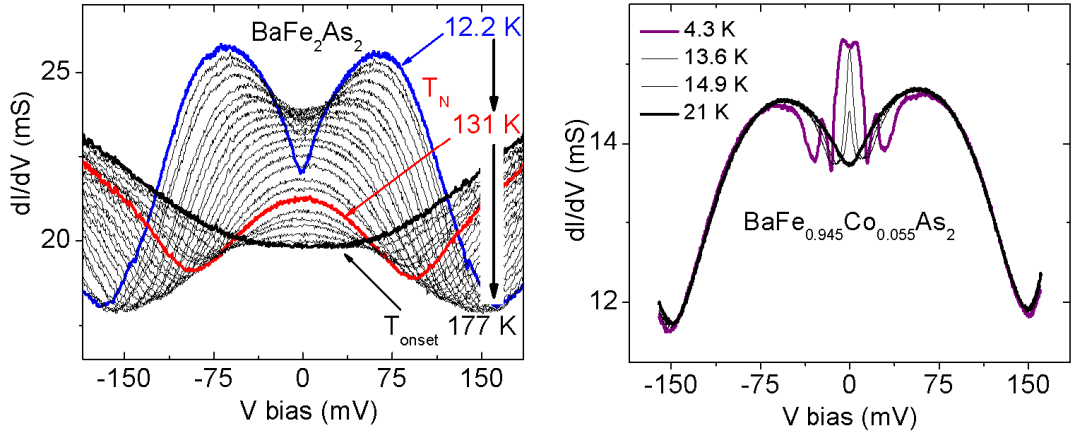

Figure 1. PCS conductance spectra are shown over broad voltage and temperature ranges. The left panel spectra are taken on the parent compound BaFe$_2$As$_2$. Note the onset temperature of the GLF excess conductance is ~177 K, well above T$_N$ (red). The right panel shows conductance spectra for the superconducting compound BaFe$_{0.945}$Co$_{0.055}$As$_2$. Note the GLF remains at high bias, and Andreev reflection is observed below T$_c$ [After Ref. 36].

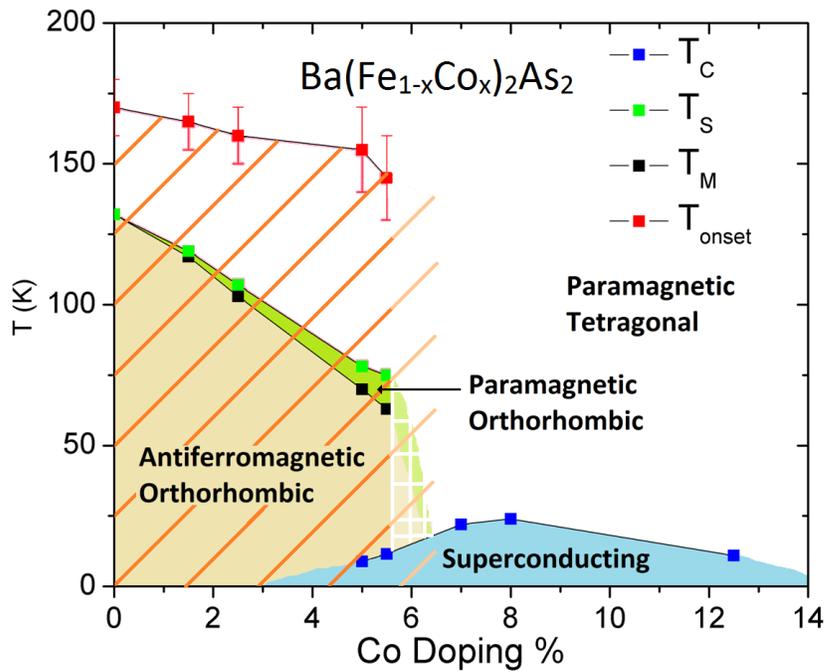

Figure 2: The red squared denote the onset of the GLF, above the antiferromagnetic and orthorhombic transitions. The GLF remains to the lowest temperatures measured, within the pink boarder on the under-doped side of the phase diagram [After Ref. 36].